\documentstyle[aps,prl,multicol]{revtex}
\begin{document}
\draft
\title{Trapped $^{6}$Li : A high T$_{c}$ superfluid ?}
\author{R. Combescot}
\address{Laboratoire de Physique Statistique,
 Ecole Normale Sup\'erieure*,
24 rue Lhomond, 75231 Paris Cedex 05, France}
\date{Received \today}
\maketitle

\begin{abstract}
We consider the effect of the indirect interaction due to the exchange of 
density fluctuations on the critical temperature of superfluid $^{6}$Li . 
We obtain the strong coupling equation giving this critical temperature. 
This equation is solved approximately by retaining the same set of 
diagrams as in the paramagnon model. We show that, near the 
instability threshold, the attractive interaction due to density fluctuations 
gives rise to a strong increase in the critical temperature, providing a 
clear signature of the existence of fluctuation induced interactions.
\end{abstract}

\begin{multicols}{2}
The recent quite impressive progress in obtaining ultracold atomic 
gases has open the way to the discovery of a large number of new 
superfluids. Already most alkali Bose gases have been shown to  
undergo Bose Einstein condensation. Even if superfluidity has not been 
yet demonstrated explicitely in experiments, they are firmly believed to 
be superfluids because they have already been seen to display phase 
coherence. Moreover the search for the transition of trapped Fermi 
gases toward a BCS type superfluid is now actively considered 
\cite{stoofal}, as the possibility of experimental 
observation is quite realistic.

A high critical temperature will be obtained for a strong attractive 
effective interaction between fermions. Since at low temperature 
scattering is essentially s-wave, a large negative scattering length is 
most favorable. Therefore spin polarized $^{6}$Li appears as a strong 
candidate since  its scattering length is found experimentally 
to be very large \cite {abra} $a$ = - 1140 $\AA $ . In 
this case pairing would occur between two different  hyperfine states 
\cite{stoofal}, which play the same role as spin states in  standard 
BCS theory. This is the situation we will consider and we will assume 
the most favorable case where the number of atoms is the same in the 
two hyperfine states. It is naturally important to assess as well as 
possible the value of the critical temperature. Experiments are presently  
performed in magnetic atomic traps leading to a harmonic potential and 
calculations should be done for this geometry \cite{castin}. Or one can 
also consider making use of an optical trap. Here we will restrict 
ourselves to the simpler situation of fermions confined in a box with 
total density $n= k_{F}^{3}/3\pi ^{2}$.

The high density regime is of particular interest since
it corresponds clearly to higher critical temperature and the 
superfluid will be more accessible experimentally in this regime.
However there is also a 
deep theoretical interest to investigate this domain. Indeed the high 
density regime is bounded by an instability which occurs for 
a coupling constant $ \lambda = 2 k_{F} | a | / \pi \geq 1 $ . 
Beyond this limit the compressibility becomes negative because of the 
strong effective attraction between atoms in different hyperfine 
states. In the vicinity of this instability 
$ \lambda \lesssim 1 $, the compressibility becomes high and 
density fluctuations occur easily. This leads to the possibility of an 
attractive interaction between fermions through the exchange of 
density fluctuations, in a way completely analogous to the phonon 
exchange mechanism of standard superconductivity. Qualitatively we 
expect this mechanism to add up to the direct attractive interaction and 
to lead to an increase in the critical temperature.

Independently of this increase, this situation is extremely interesting 
because it is quite analogous to what is believed to happen in many 
other condensed matter systems. High $T_{c}$ superconductors and 
the newly discovered low temperature superconductor $ Sr _{2}RuO_{4} $
are well-known examples. However the best case is probably liquid 
$^{3}$He \cite{vowo} which is not so far from having a 
ferromagnetic instability. It has been proposed that strong spin 
fluctuations, the so-called paramagnons, exist in this liquid 
\cite{doen,besc} and play an important role in the physics.
In particular the attractive 
interaction, leading to the Cooper pairs formed in the superfluid, has 
been attributed to paramagnon exchange. A strong qualitative support 
for this picture is the existence  of the A  phase of superfluid $^{3}$He 
at high pressure, which has been explained  \cite{anbr} by feedback 
effects of its specific structure on the  paramagnon  propagator. 
However it is not clear at all that the paramagnon model can  provide a  
quantitative description of the properties of the liquid, and in particular  
can  account  for the observed values of the critical temperature 
\cite{leva}.

In this respect the situation in $^{3}$He is difficult, since one does not  
have  a  precise knowledge of the instantaneous part of the pairing 
interaction.  Moreover the  parameter $\bar{I}$ involved in the 
paramagnon description varies in a  quite  restricted range in the vicinity 
of the instability limit $\bar{I}$ = 1, when the pressure of the liquid is 
varied over the full range available in the phase  diagram.  By contrast 
the situation potentially offered by superfluid $^{6}$Li gas is much 
more agreable : the instantaneous interaction can be directly linked to 
the  diffusion  length, which is fairly precisely known. Moreover the 
possibility of varying to a  large extent the density allows to change the 
coupling constant at will. This offers a stringent coherence check of the 
theory since experiment will be able to verify the general qualitative 
behaviour of the model. In  this way we can hope to have a definite answer 
to the question : is the paramagnon  model a proper description or not ?

The problem of the critical temperature for a BCS superfluid in a dilute 
Fermi gas has been investigated by Gorkov and Melik-Barkhudarov 
\cite{gork}, following the work \cite{gal} of Galitskii. It has been 
recently considered in the more general context of dilute atomic gases 
\cite{stoofal} by Stoof et al. who found a typical value of 40 nK for a 
density $ n = 10 ^{12} cm ^{-3} $. This temperature is quite within 
reach experimentally. Our purpose is to extend these works in the 
high density regime. Naturally our result will reduce to the proper 
one in the dilute limit. 

The critical temperature is obtained \cite{gork} by writing that the 
vertex part in the normal state diverges, which is expected to 
occur first for zero total momentum and energy 
of the pair. The corresponding vertex part $ \Gamma _{p,p'}$ is 
related to the irreducible vertex $ \bar{\Gamma}_{p,p'} $ 
in the particle-particle channel by :
\begin{eqnarray}
\Gamma _{p,p'}=\bar{\Gamma}_{p,p'} - 
T \sum_{k} \bar{\Gamma}_{p,k}D_{k} \Gamma _{k,p'}
\label{eq1}
\end{eqnarray}
where $ \sum_{k}$ is for $ (2\pi )^{-3} \sum_{n} \int d \bf k $. 
We have set $ D_{k} = G(k) G(-k) $ where $ G(k)$ is the full Green's 
function and $k=({\bf k},\omega _{n})$ is a four-momentum. 
The summation runs over the wavector ${\bf k}$ and the 
Matsubara frequency $\omega _{n}$ = $ ( 2n + 1 ) \pi T $. 

We split the irreducible vertex into the bare interaction 
$ U _{ \bf{p}, \bf{p}' } $ 
and all the contributions $ \Gamma ^{\star}_{p,p'}$ which are higher 
order in the interaction :  $ \bar{\Gamma} _{p,p'}$ = $U _{ \bf{p}, 
\bf{p}'}$ + $ \Gamma ^{\star}_{p,p'}$ . Then, following Galitskii 
\cite{gal}, we eliminate the interaction U in favor of the vertex in the 
dilute limit corresponding physically to two atoms scattering in 
vacuum. In this limit, $ 
\bar{\Gamma}$ reduces to U, and $ G(k)$ becomes the free particule 
Green's function  $ ( i \omega _{n}- \epsilon  _{k}) ^{-1}$ with $ 
\epsilon  _{k} = k ^{2}/2m $. In
contrast to Ref. \cite{gork} , we have taken for convenience the 
chemical potential equal to zero in this limit. Let us call $ \Gamma ^{T} 
_{\bf{p}, \bf{p}'} $ the vertex in this limit. From Eq.(1) it 
satisfies :
\begin{eqnarray}
\Gamma ^{T} _{{\bf p}, {\bf p}'} = U _{ {\bf p}, {\bf p}'} - 
\frac{T}{(2\pi )^{3}} \int d {\bf k} \: U _{ {\bf p}, {\bf k}} \sum_{n} 
D ^{0}_{k} \Gamma ^{T} _{{\bf k}, {\bf p}'}
\label{eq2}
\end{eqnarray}
where $ D ^{0}_{k} = ( \omega _{n} ^{2} + \epsilon  _{k}^{2} ) ^{-
1}$. Since we will deal with fairly small temperature, we can take 
the  T $ \rightarrow 0 $ limit where $ \Gamma ^{T} 
_{{\bf p}, {\bf p}'} $ reduces to the vertex $ \Gamma ^{0} _{ {\bf 
p}, {\bf p}'} $ for two scattering atoms evaluated at zero energy. This 
vertex can be explicitely expressed \cite{fw} in terms of the scattering 
amplitude $ f ({\bf p}, {\bf p}') $ corresponding to the scattering 
potential U. Since the atomic potential has a 
very short range compared to all other lengths
involved in the problem, the typical wavevector for a
change in $ U ( {\bf p}, {\bf p}') $ is very large compared to the
wavevectors we have to deal with and similarly for
$ \Gamma ^{0} _{{\bf p}, {\bf p}'} $. Hence we can take 
in our problem $ \Gamma ^{0} _{{\bf p},{\bf p}'} $ equal 
to its $ \bf{p}= \bf{p}' = 0 $ limit, which is given \cite{fw} 
by $ \Gamma ^{0} = 4 \pi a/m $ in terms of the scattering length.
This can be checked explicitely in the case of a
separable potential or for a pseudopotential \cite{hcct}.

Since $ \Gamma _{p,p'} $ diverges at $ T _{c} $, the first term in the 
r.h.s. of Eq.(1) is negligible. In the resulting integral equation, 
$p'$ appears as a free parameter which can be omitted.
Writing $ \Gamma _{p}$ instead of $ \Gamma _{p,p'}$, 
we obtain that $ T =  T _{c} $ when:
\begin{eqnarray}
\Gamma _{p}= - T \sum_{k} [ U _{ 
{\bf p}, {\bf k}} + \Gamma ^{\star}_{p,k} ]D_{k} \Gamma _{k}
\label{eq3}
\end{eqnarray}
is satisfied. Note that the 
effective interaction $ \Gamma ^{\star}_{p,k} $ will be frequency 
dependent. Therefore we have also to retain the frequency dependence 
of $ \Gamma _{k} $. In other words the order parameter in the 
superfluid phase will have a frequency dependence, which corresponds 
to a strong coupling situation. We eliminate $U$ from Eq.(3) by 
premultiplying it by $ \delta _{p',p} - T \Gamma ^{0}_{ {\bf p}', {\bf p}}
D ^{0}_{p}$ and summing over $p$. Making use of Eq.(2) this leads to:
\begin{eqnarray}
\Gamma _{p}= - \Gamma ^{0} T \sum_{k}[ D_{k} - D ^{0}_{k} ] 
\Gamma _{k} & & \nonumber \\
- T \sum_{k}[ \Gamma ^{\star}_{p,k} - \Gamma ^{0} T \sum_{k'} 
D ^{0}_{k'} \Gamma ^{\star}_{k',k} ] D_{k} \Gamma _{k} & &
\label{eq4}
\end{eqnarray}

Since we expect the dependence $ \Gamma _{p} $ to be fairly slow,
over a typical scale $k _{F}$, with respect to the wavevector {\bf p},
we will neglect it in the following.
For coherence $ \Gamma ^{\star}_{p,k} $ in 
the r.h.s. of Eq.(4) will be evaluated for $ | {\bf p} | = k _{F} $.
Similarly we can evaluate the self-energy
$ \Sigma (k) $ at the Fermi surface. In terms of the renormalization 
function $ Z _{n}  $ defined by  $ 
\Sigma (k _{F}, i \omega _{n}) - \Sigma (k _{F}, 0) = i \omega _{n} 
( 1 - Z _{n}) $ this leads to
$ D_{k} = (  \omega _{n} ^{2} Z _{n} ^{2}+ \xi  _{k}^{2} ) ^{-1} $
with $ \xi _{k} = \epsilon _{k} - E _{F}$. This procedure
is the standard one for classical superconductors, fully justified 
by the existence of a small energy scale which is the phonon energy. 
Here to the contrary we have the single energy scale $E _{F}$, and the 
quality of this set of approximations is less obvious. Nevertheless, 
since one does not expect any drastic effect to arise from this {\bf k} 
dependence and that this procedure is consistent with the 
paramagnons approach, we will work with this approximate scheme.

Once $ \Gamma _{p} $ has no {\bf p} dependence, integrations over 
momentum can be performed in Eq.(4). In the first term, we write $ 
D_{k} - D ^{0}_{k} = [ D_{k} - D ^{1}_{k} ] + [ D ^{1}_{k} - D ^{0}_{k} ] $  
where $ D ^{1}_{k} = (  \omega _{n} ^{2} + \xi  _{k}^{2} ) ^{-1} $.  
The contribution from $ D ^{1}_{k} - D ^{0}_{k} $ can be integrated 
exactly. For $ D_{k} - D ^{1}_{k} $ we take into account, consistently
with our above approximation, that $ \omega _{n}$ is small.
In this way we obtain :
\begin{eqnarray}
\frac{1}{(2\pi )^{3}} \int d {\bf k} \:  [D_{k} - 
D ^{0}_{k}] =  \frac{\pi N _{f}}{| \omega _{n}| } 
[ C _{n} ^{W} - C _{n} ^{M} ]
\label{eq5}
\end{eqnarray}
where $ N _{f} = m k _{F}/2 \pi ^{2} $ is the density of states at the 
Fermi surface. We have set $ C _{n} ^{M} = 1 - Z _{n} ^{-1} $ 
and $ C _{n} ^{W} = [\sqrt{1+(1+w _{n} ^{2}) ^{1/2}} - \sqrt{|w _{n}|}]
/ \sqrt{2} $ with $ w _{n} = \omega _{n} / E _{F} $. One can 
check that the term $ C _{n} ^{W} $ gives in Eq.(4)  a contribution 
$\Gamma ^{0} N _{f} \ln (8 e ^{C-2}E _{F} / \pi  T) $ (C is the 
Euler constant). If only this term is retained one obtains 
$ T _{c}/ E _{F} = 8 e ^{C-2} / \pi  \exp (-1/ \lambda ) $ used in 
\cite{stoofal}. The $ C _{n} ^{M} $ term will give a decrease of the 
critical temperature due to mass renormalization and lifetime effects. 

We turn now to the irreducible vertex $ \Gamma ^{\star}_{k,k'}$. 
As we have indicated we are mostly 
interested in the contribution of density fluctuations to this vertex, and 
here we will handle it by retaining the same set of diagrams as in 
paramagnon theory. Actually since the attractive interaction acts only 
between atoms with different hyperfine states, we are exactly in the 
same situation as for paramagnons where interaction takes place only 
between different spins. The only qualitative difference is that the 
interaction is repulsive in paramagnon theory, leading to a positive 
dimensionless coupling constant $ \bar{I}= N _{f} I $, while the 
attractive interaction between $^{6}$Li correspond to a negative $ 
\bar{I} $. Another important difference is that it would be inaccurate to 
retain only the bare interaction for all the elementary vertices in the 
paramagnon diagrams. Indeed we know that we have a large scattering 
length. This is obtained quantitatively by summing up the ladder 
diagrams for two scattering atoms, as it is clear from Eq.(2). Obviously 
we have to do a similar summation in the paramagnon diagrams, 
otherwise we will miss the dominant contribution. More precisely we 
would need to know the irreducible vertex in the particle-hole channel. 
We will assume that the dominant contribution to this vertex is given by 
the sum of the ladder diagrams. Also we will not attempt to take into 
account its energy dependence and consider that this vertex is the same 
as for two atoms in vacuum. With these hypotheses we are led to take 
the interaction $I$ of paramagnon theory equal to $\Gamma ^{0}$. 
This gives us $ \bar{I}= N _{f} \Gamma ^{0}= - \lambda $. 

Now the sum of the paramagnons diagrams ( including ladder and 
bubble diagrams ) gives \cite{leva,naka} $ \Gamma ^{\star}_{k,k'}
= N _{f} V _{\rm eff} (k - k') $ with
$N _{f} V _{\rm eff} (k) = \bar{I} ^{2} \bar{ \chi }  _{0}/(1 
- \bar{I} \bar{ \chi  }_{0}) + \bar{I} ^{3} \bar{ \chi } _{0} 
^{2}/(1 - \bar{I} ^{2} \bar{ \chi } _{0}^{2})$
and $\bar{\chi } _{0}(k) $ is the dimensionless elementary bubble 
\cite{leva} .
When the self-energy is evaluated with the corresponding set of diagrams, 
one finds \cite{leva} that it is given by the same expression as for an 
effective interaction $ N _{f} V _{Z} (k) = 
\bar{I} ^{3} \bar{ \chi } _{0} ^{2}
/(1 - \bar{I} \bar{ \chi  }_{0})+ \bar{I} ^{2} \bar{ \chi } 
_{0} / (1 - \bar{I} ^{2} \bar{ \chi } _{0}^{2})$
In agreement with our above approximation, we evaluate the self-
energy for wavevector $ k _{F}$ and for small energies. This leads to:
\begin{eqnarray}
(2n+1) ( Z _{n} - 1 ) = \bar{V} _{Z}(0) + 2 \sum 
^{n}_{p=1}\label{n}\bar{V} _{Z}(2 \pi pT)
\label{eq6}
\end{eqnarray}
with $ \bar{V} _{Z}(\omega _{p}) = (1/2 k^{2} _{F}) \int _{0} ^{2}qdq N _{f} 
V _{Z} (q,\omega _{p})$. This $ \bar{V} _{Z}(\omega _{p}) 
$ corresponds to the average of the interaction for scattering of an atom 
at the Fermi surface from $ {\bf  k}$ to $ {\bf  k}'$ with wavevector 
transfer $ q = 2 k _{F} \sin ( \theta /2 ) $ where $ \theta $ is 
the angle between $ {\bf  k}$ and $ {\bf  k}'$. This angular average 
has to be performed numerically, in order to obtain $ \bar{V} _{Z} $ . 
As expected $Z _{n} > 1$. The maximum is at zero energy, with a fairly 
long tail at high energy. Naturally $ Z _{n} $ increases when 
$ |\bar{I}| \rightarrow 1 $ and it diverges at zero energy in this limit.

In the term $ \Gamma ^{\star}_{p,k} D_{k} \Gamma _{k} $ in Eq.(4), the 
only strong dependence on {\bf  k}, for low energies, comes from 
$D_{k}$ which forces $ k \approx k _{F}$. So we integrate over $ \xi 
_{k}$ as above.
Setting $  \Gamma _{n} = \Gamma ( \omega _{n})$, this leads us to:
\begin{eqnarray}
T \sum_{k} \Gamma ^{\star}_{p,k} D_{k} \Gamma _{k} = - \sum _{m} 
\frac{ \pi T \: \Gamma _{m}}{| \omega _{m} | Z _{m}}  \bar{V} _{\rm 
eff}(\omega _{n} - \omega _{m}) 
\label{eq7}
\end{eqnarray}
where as above $ \bar{V} _{\rm eff}(\omega _{p}) = - (1/2 k^{2} _{F})
 \int _{0} ^{2}qdq N _{f} V _{\rm eff} (q,\omega _{p})$. 
Finally the contribution from the last term in Eq.(4) with 
the double summation is somewhat more complicated to evaluate. 
Nevertheless after a double integration, 
performed numerically without problems, which provides a function
$V _{c}( \omega _{n})$, it can be written as:
\begin{eqnarray}
T ^{2} \sum_{k,k'} D ^{0}_{k'} \Gamma ^{\star}_{k',k} D_{k} 
\Gamma _{k} = N _{f} \sum _{m} \frac{ \pi T \: \Gamma _{m}}{| \omega _{m} 
| Z _{m}} V _{c}( \omega _{m})
\label{eq8}
\end{eqnarray}

Finally Eq.(4) can be cast as:
\begin{eqnarray}
\Gamma _{n} = \sum _{m} \frac{ \pi T}{| \omega _{m} | } \lambda  C 
_{m} \Gamma _{m} + \frac{ \pi T}{| \omega _{m} | Z _{m}} \bar{V} 
_{\rm eff}( \omega _{n}- \omega _{m}) \Gamma _{m} 
\label{eq9}
\end{eqnarray}
where we have set $C _{n} = C _{n} ^{W} - C _{n} ^{M} - C _{n} 
^{V}$ with $C _{n} ^{V} = V _{c}( \omega _{n}) / Z _{n}$.
Hence $T _{c}$ is the highest temperature for which the matrix 
corresponding to the r.h.s. of Eq.(9) has an eigenvalue equal to 1.
Eq.(9) is very similar to the one obtained from 
Eliashberg equations for a strongly coupled superconductor. The most 
noticeable difference is the term $C _{n} ^{V}$ produced by the 
interaction through fluctuation exchange when one replaces the bare 
potential in terms of the scattering length. We note also that, in the 
dilute limit, we recover the result of Gorkov and Melik-Barkhudarov 
\cite{gork}.

Let us now consider the physical effect of ${V} _{\rm eff}$. It is more 
convenient to decompose it \cite{anbr,naka} as
$N _{f} V _{\rm eff} = (3/2) \bar{I} ^{2} \bar{ \chi }  
_{0}/(1 - \bar{I} \bar{ \chi  }_{0}) - (1/2) \bar{I} 
^{2} \bar{ \chi } _{0}/(1 + \bar{I} \bar{ \chi } _{0})$.
The first term is due to spin fluctuations and the second one 
to density fluctuations. As pointed out by Berk and 
Schrieffer \cite{besc} the first term gives always a repulsion
which increases strongly in the 
vicinity of the ferromagnetic instability $ \bar{I} \rightarrow 1 $.
However in the range $ \bar{I} < 0 $ which 
is of interest to us, the density fluctuations attraction can take over the 
spin fluctuations repulsion. This is what happens in the limit $ \bar{I} 
\rightarrow -1 $ where the gas becomes unstable. Then in complete 
analogy with the paramagnon case, the effective interaction at the Fermi 
surface $\bar{V} _{\rm eff}( \omega _{n})$ diverges for zero 
frequency. For nonzero frequency it remains finite. In particular at high 
energy where $\bar{ \chi }  _{0}$ gets small, spin fluctuations 
dominate, $ N _{f} V _{\rm eff} \approx \bar{I} ^{2} \bar{ \chi }_{0} 
$ and the overall interaction is repulsive. So for  $\bar{I}$ near -1, the 
effective interaction is attractive at low frequency and has a repulsive 
high energy tail. But the attractive part exists only for $\bar{I} \lesssim
-0.6$. For lower $\bar{I}$ fluctuations are repulsive for all energies. 

The divergence of $\bar{V} _{\rm eff}$ for $ \bar{I} \rightarrow -1 $ 
may raise hope that in this limit $T _{c}$ is going to be very large. 
However it is known, in the case of strongly coupled superconductor 
as well as for \cite{leva} $^{3}$He in paramagnon theory, that this 
increase is countered by the corresponding increase of $Z _{n}$, 
physically due to mass renormalization and lifetime effects. The net 
result of these opposite effects has to be found numerically, so we turn 
to our numerical results from Eq.(9) for $T _{c}$.
They are given in Fig.1.

Surprisingly, instead of a regular rise of $T _{c}/ E _{F}$ as a 
function of $ \lambda = - \bar{I} $, we find two regimes. Up to $ 
\lambda \approx 0.4 $ we have a regular and strong increase (note that 
for $ \lambda = 0.4 $ our result is markedly below the extrapolation of 
the result of Ref. \cite{gork}, which would give $T _{c}/ E _{F}$ = 
0.023). We have then a saturation and even a slight decrease up to $ 
\lambda \approx 0.6 $. We attribute this effect to the increase of $Z 
_{n}$  which overcompensates the increase with $ \lambda $ of the 
direct attractive interaction. Then, starting for $ \lambda \approx 0.6 $ 
we obtain another strong rise of $T _{c}/ E _{F}$. This second 
regime is clearly due to the increasing contribution of the attractive 
interaction due density fluctuations.
$T _{c}/ E _{F}$ grows up to a maximum 0.025  found for 
$ \lambda \simeq 0.98 $ (corresponding to $T _{c} $ = 190 nK). We 
stress that, if observed, this critical temperature would be the highest 
(relative to $ E _{F} $) among BCS superfluids, since for standard 
superconductors as well as superfluid $ ^{3}$He this ratio is of order 
10$ ^{-3}$ whereas for high $T _{c}$ superconductors (if they are 
BCS) it reaches at best 10$ ^{-2}$ . Then, when $ \lambda $ increases 
further, $T _{c}$ decreases gently in the vicinity of $ \lambda = 1 $. 
This effect is clearly due to the increase of $Z _{n}$ .
However the maximum of $T _{c}$ is obtained for $ \lambda $ so close 
to 1 that this decrease is probably unobservable experimentally. 

The most interesting feature of these results is that the existence of an 
indirect attractive interaction due to density fluctuations exchange gives 
a qualitative signature in $T _{c}( \lambda ) $. Although we may 
wonder about the quantitative validity of our results, due to the 
approximations we have made, it is reasonable to believe 
that the qualitative rise in $T _{c}( \lambda ) $ will survive. Its 
observation would be a strong indication of the importance of 
fluctuation exchange in the effective interaction in $^{6}$Li. And 
indirectly it would also bring some support to the existence of similar 
mechanisms in other BCS superfluids. Finally in this regime we 
expect, in the superfluid phase, deviations from standard weak 
coupling BCS theory and feedback effects. In conclusion we have shown 
that, for $^{6}$Li near the instability threshold, indirect 
attractive interaction due to density fluctuations exchange 
can lead to rather high critical temperature, with a clear 
signature for its dependence as a function of the coupling constant. The 
observation of this effect would be very interesting, as a clear example 
of collective mode induced superfluidity.

We are very grateful to Y. Castin, C. Cohen-Tannoudji, J. Dalibard, 
W. Krauth, M.O. Mewes and C. Salomon for very stimulating 
discussions.

* Laboratoire associ\'e au Centre National
de la Recherche Scientifique et aux Universit\'es Paris 6 et Paris 7.

\begin{figure}
\caption{$T _{c}/ E _{F}$ as a function of the coupling constant $ \lambda $.
The dots correspond to actual calculations, the line is just 
a guide for the eye.}
\label{Fig1} 
\end{figure}

\end{multicols}
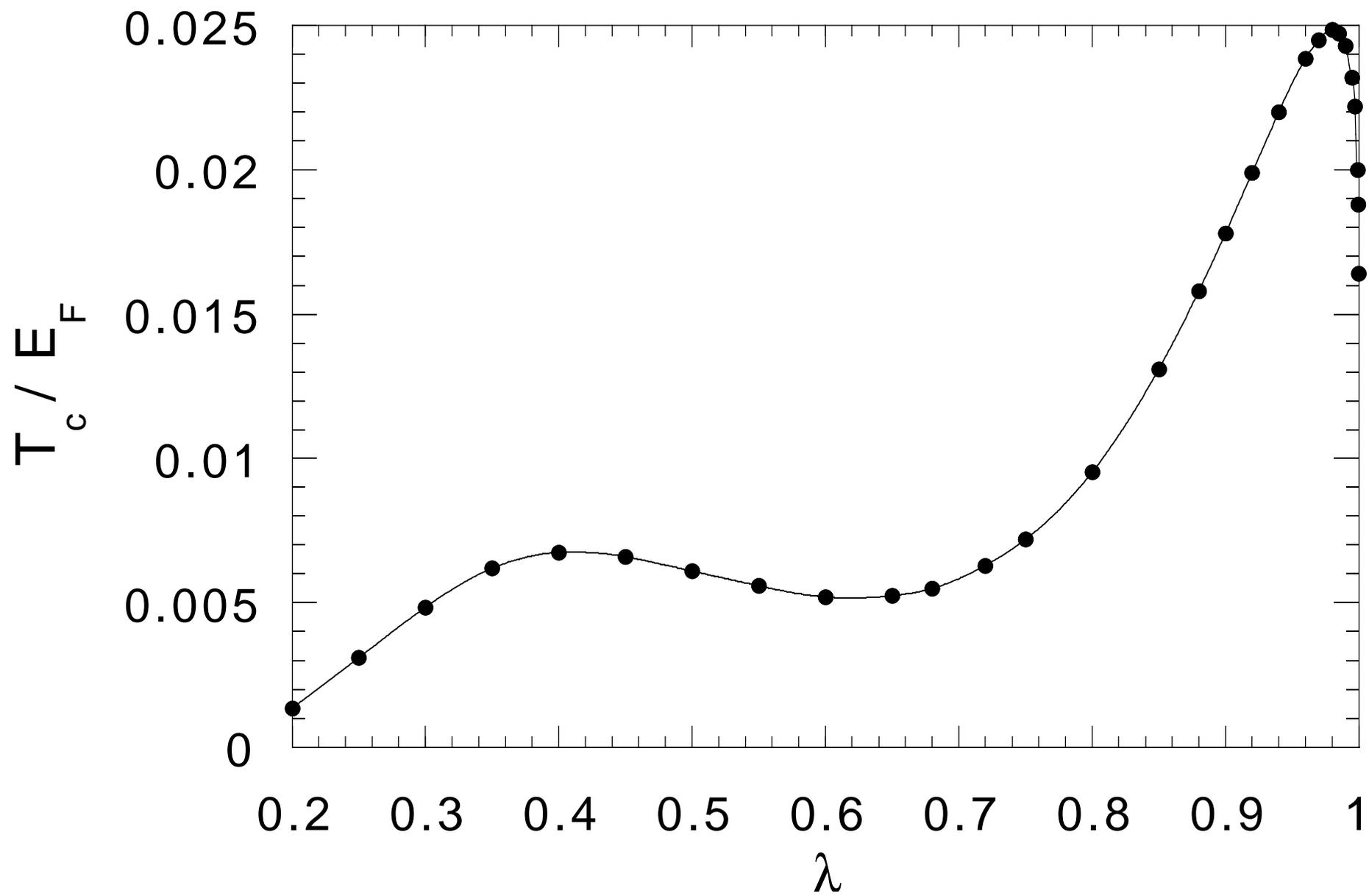
\end{document}